\documentclass[a4paper]{jpconf}
\usepackage{graphicx}
\begin{document}
\title{Collectivity of the electromagnetic transitions in near-threshold resonances}

\author{M. P{\l}oszajczak}

\address{Grand Acc\'el\'erateur National d'Ions Lourds (GANIL), CEA/DSM - CNRS/IN2P3,
BP 55027, F-14076 Caen Cedex, France}

\ead{ploszajczak@ganil.fr}

\author{J. Oko{\l}owicz}

\address{Institute of Nuclear Physics, Polish Academy of Sciences, Radzikowskiego 152, PL-31342 Krak{\'o}w, Poland}

\ead{Jacek.Okolowicz@ifj.edu.pl}

\begin{abstract}
Mixing of the shell model (SM) eigenstates due to the coupling via the common decay channel leads in many cases to the formation of a collective eigenstate which carries many features of the nearby decay channel. This generic mechanism in open quantum systems explains  the phenomenological Ikeda diagram and generalizes it for various clusters/correlations in the vicinity of the respective cluster decay thresholds. The near-threshold collectivization of the SM states may also influence their electromagnetic decays. We discuss this phenomenon on the example of B(E$\lambda$) decays of near-threshold 2$^+$ states in $^{14}$C. 
\end{abstract}

\section{Introduction}

Since the beginning of century, the low-energy nuclear theory evolves rapidly. New innovative strategies of solving nuclear many-body problem have been proposed. Regarding nuclear many-body approaches, the field has gone from the Green's function Monte-Carlo calculations to the coupled cluster formalism to the no-core shell model to the in-medium similarity renormalization group, as well as other techniques.  

  Besides nuclear structure, there have also been great advances in nuclear reaction theory, namely the problem of how to include the continuum in the theory of nuclear properties.  This has been aided by the resonating group method and the Berggren basis for including the states in the continuum. This work has led, for example, to the no-core shell model coupled with the resonating-group method \cite{navratil08,baroni13} and the no-core Gamow shell model \cite{mic02,betan02,JPG_GSM_review} for calculating continuum properties, such as resonances and scattering configurations. Further development of the shell model embedded in the continuum (SMEC) \cite{benn00,rot05,Oko03} allowed for a microscopic description of near-threshold clustering and two-proton radioactivity using realistic SM interactions. 

Gamow shell model (GSM) completes the nuclear SM and offers a fully symmetric treatment of bound, resonance and scattering states which preserves the unitarity of many-body calculation at around the particle emission threshold. The mathematical setting of the GSM follows directly from the formulation of quantum mechanics in the rigged Hilbert space \cite{rhs}. 

The deeper understanding of nuclear properties which is provided by SM for open quantum systems defines new territory for spectroscopic studies which extends from drip lines to the region of stable nuclei for states in the vicinity and above the first particle emission threshold. Systematic studies in this broad region of masses and excitation energies will extend and complete our knowledge of atomic nucleus at the edge of stability.  

The challenges for nuclear theory cannot be separated from new experimental discoveries and the significant efforts to improve nuclear experimentation techniques. From the exciting new research perspectives for nuclear theory, we will discuss here the open quantum system perspective on atomic nucleus, and in particular, the role of external configuration mixing of SM states through the continuum as a mechanism of the collectivization of near-threshold states.

\section{Evolution of the nuclear theory paradigms}

SM provided foundation of modern nuclear theory and helped to comprehend large amount of data on nuclear levels, moments, collective excitations, and various kinds of decays. SM describes the nucleus as a closed quantum system (CQS): nucleons occupy bound, localized, single-particle orbits, and are isolated from the environment of scattering states. 

First open quantum system formulation of the nuclear SM, respecting unitarity at the particle emission threshold(s), has been achieved in GSM \cite{mic02,betan02,JPG_GSM_review}. The many-body states in GSM are given by the linear combination of Slater determinants defined in the Berggren ensemble \cite{berggren_1968} of single-particle states which includes bound states, resonances and complex energy scattering states along the respective contours in the complex $k$-plane. Reaction channels are not explicitly identified so GSM in the Slater determinant representation is the tool for spectroscopic studies of bound and unbound states and their decays. 

For the description of scattering properties and reactions, the entrance and exit reaction channels have to be identified. This can be achieved by expressing GSM wave functions in a complete basis of the reaction channels. This coupled-channel representation of the GSM (CC-GSM) has been recently applied for various observables, such as the excitation function, the proton/neutron/deuteron elastic/inelastic differential cross-sections \cite{PRC_GSM_CC_Yannen,Mercenne16,michel2019}, and the low-energy proton/neutron radiative capture cross-sections \cite{PRC_GSM_CC_Kevin,Dong17}. One should stress that channels in CC-GSM are build by GSM wave functions which respect the unitarity at the decay threshold of each cluster subsystem. 

One could ask, why should we care about the continuum couplings in spectroscopic studies? At the limit of nuclear stability with respect to the particle emission, {\em i.e.} in the vicinity of drip-lines or near the particle emission threshold in well-bound stable nuclei, nuclear states belong to the multidimensional network of  states interconnected via the coupling to decay channels and scattering states. This network of open quantum system (OQS) eigenstates spans the 3D-lattice in the space of proton numbers, neutron numbers, and excitation energy. Traditional nuclear structure theory describes nucleus as the CQS which is separated from the continuum of scattering states and decay channels. States of the CQS form the 1D-chain and, hence, the ensemble of states in all nuclei forms a forrest of separate 1D networks. Properties of both quantum and classical networks depend on the lattice dimension and on the nature of interaction between sites (configurations) of the lattice.  Therefore, the CQS description of weakly bound or unbound states yields a crooked-mirror image of their properties. Indeed, new phenomena which are unknown in CQSs, such as the coalescence of eigenfunction/eigenvalues, the segregation of time scales, the near-threshold collectivity and clustering, the multichannel effects in cross-section and shell occupancies, the continuum-induced breaking of mirror symmetry and isospin symmetry, the violation of orthogonal invariance and channel equivalence, etc., are expected in weakly bound or unbound nuclear states. 

The OQS perspective on nuclear properties changes also objective of the nuclear experimentation  which should aim not only at the understanding of properties of individual states and their decays in a given nucleus, but should provide the understanding of the ensemble of states and their mutual connections in the neighbouring nuclei to disclose the basic features of the domains of correlated states in different regions of excitation energy and proton/neutron numbers. 

Many near-threshold effects should be studied experimentally, {\em e.g.} what are the $\gamma$-selection rules in the continuum for in- and out-of-band transitions? The studies in the dipolar anions \cite{Kevin_dipolar_anions} have demonstrated that states of the rotational bands above the dissociation threshold are strongly $K$-mixed whereas below this threshold all states have $K$=0. Similarly, what is the nature of near-threshold $\gamma$-transitions? Are they strongly influenced by the collectivization of SM states via the coupling to the particle emission threshold(s)?

Understanding the nature of pairing correlations in weakly-bound states and in the continuum is yet another great challenge. It has been shown \cite{Luo} that the $T=1$ pairing correlations are strongly modified by the $T=0$ neutron-proton continuum coupling in odd-$Z$ chain of isotopes if one-neutron $S_n$ and two-neutron $S_{2n}$ separation energies tend to zero. In this limit, an anti-odd-even staggering of continuum coupling energy correction leads to a significant decrease of the odd-even staggering of binding energies and hence a decrease of the pairing gap while the pair amplitudes remain unchanged. Hence the blocking mechanism weakens and a gradual transition towards the gapless superconductivity appears. 
 
Occupation of the single-particle shells can be modified in the vicinity of the particle emission threshold. GSM studies \cite{michel07} have shown that one-nucleon spectroscopic factor shows an anomalous behaviour near neutral particle emission threshold with a characteristic dependence $(-S_n)^{\ell-1/2}$ below the threshold and $(-S_n)^{\ell+1/2}$ above the threshold, in a complete analogy with the Wigner threshold phenomenon for reaction cross-sections. This unusual dependence of spectroscopic factors is a result of an interplay between discrete resonant states and non-resonant continuum in the many-body wave function. It is also a direct manifestation of the unitarity which is violated in SM at each consecutive particle emission threshold. 

Near the charged-particle emission threshold, there is no cusp behavior in the spectroscopic factor for low angular momenta $\ell$ \cite{michel10}. This shows that the effective correlations among neutrons and protons which determine occupancies of single-particle shells, act differently depending on whether the state is in the proximity of neutral- or charged-particle threshold. This finding has far going consequences for the nature of many-body states at the proton and neutron driplines, for the average correlations that protons (neutrons) experience in proton-rich (neutron-rich) matter, or for the structure of mirror states.

\section{Emergence of near-threshold collectivization and clustering}

Ikeda et al. \cite{ikeda} observed that $\alpha$-cluster states can be found in the proximity of $\alpha$-particle decay thresholds in light nuclei. This finding cannot be a consequence of any specific feature of nuclear interaction because then the nature of both nucleon-nucleon correlations and clustering in near-threshold states would appear at random. Hence, the origin of cluster (correlated) states in the proximity of cluster-decay thresholds must be more general. Based on the results of SMEC, it has been conjectured \cite{cluster} that the interplay between internal configuration mixing by interactions and external configuration mixing via decay channels leads to a new kind of near-threshold collectivity.  Following this conjecture, the Hoyle resonance  close to the $\alpha$-particle emission threshold in $^{12}$C should carry an imprint of the [$^8$Be$\oplus\alpha$] decay channel, whereas the $1/2_1^+$ resonance of $^{17}$O, well above the neutron emission threshold and in the vicinity of the $\alpha$-particle emission threshold should carry the imprint of [$^{13}$C$\oplus\alpha$] decay channel and not of the [$^{16}$O$\oplus n$] channel. 
Similarly, the ground state of $^{11}$Li should resemble the [$^{9}$Li$\oplus$2n] configuration of the nearest 2n-emission threshold rather than the [$^{10}$Li$\oplus$n] configuration, and the collectivization of $1/2_6^+$ neutron resonance in $^9$Li should be caused primarily by the proximity of [$^8$He$\oplus$p] decay channel \cite{lee}. 

Numerous SMEC studies have shown that proximity of the branch point singularity at the particle emission threshold induces collective mixing of SM eigenstates, in which the essential role is played by a single 'aligned' eigenstate of the OQS Hamiltonian which carries many characteristics of the nearby decay channel. This state is a superposition of SM eigenstates having the same quantum numbers. The point of the strongest collectivity, {\em i.e.} the centroid of the opportunity energy window, is determined by an interplay between the competing forces of repulsion (the Coulomb and centrifugal interactions) and attraction (the continuum coupling interaction). For higher angular momenta $\ell$ and/or for charged particle decay channels, the extremum of the correlation energy is shifted above the threshold. Consequently, as the Coulomb barrier increases, the collectivization due to the coupling to charged-particle decay channels becomes weaker and disappears in heavier nuclei. 

This generic phenomenon in OQSs explains why so many states, both on and off the nucleosynthesis path, exist 'fortuitously' close to open channels. In this context, one should mention that even though bound multineutrons ({\em e.g.} the tetraneutron) are incompatible with the present understanding of nuclear forces \cite{fossez17}, nevertheless the multineutron near-threshold correlations may appear and could be seen as a dynamical effects or pseudo-resonances in reaction observables. In the next section, we will show that the collectivization of the SM eigenstates may have also a noticeable effect on electromagnetic transitions and nuclear moments in weakly bound and unbound states.

\section{Electromagnetic transitions in near-threshold resonances of $^{14}$C}

The calculations are performed using the SMEC \cite{benn00,rot05,Oko03}, the recent realization of the real-energy continuum shell model \cite{Oko03,volya06}. The scattering environment is provided by one-nucleon decay channels. The Hilbert space is divided into two orthogonal subspaces ${\cal Q}_{0}$ and ${\cal Q}_{1}$ containing 0 and 1 particle in the scattering continuum, respectively. An open quantum system description of ${\cal Q}_0$  space includes couplings to the environment of decay channels through the energy-dependent effective Hamiltonian:
\begin{equation}
{\cal H}(E)=H_{{\cal Q}_0{\cal Q}_0}+W_{{\cal Q}_0{\cal Q}_0}(E),
\label{eq21}
\end{equation}
where $H_{{\cal Q}_0{\cal Q}_0}$ denotes the standard SM Hamiltonian describing the internal dynamics in the CQS approximation, and $W_{{\cal Q}_0{\cal Q}_0}(E)$:
\begin{equation}
W_{{\cal Q}_0{\cal Q}_0}(E)=H_{{\cal Q}_0{\cal Q}_1}G_{{\cal Q}_1}^{(+)}(E)H_{{\cal Q}_1{\cal Q}_0},
\label{eqop4}
\end{equation}
is the energy-dependent continuum coupling term, where $G_{{\cal Q}_1}^{(+)}(E)$ is the one-nucleon Green's function and 
${H}_{{Q}_0,{Q}_1}$, ${H}_{{Q}_1{Q}_0}$ are the coupling terms between orthogonal subspaces ${\cal Q}_{0}$ and ${\cal Q}_{1}$.
 $E$ in the above equations  stands for a scattering energy. The energy scale is settled by the lowest one-nucleon emission threshold. The coupling of internal (in ${\cal Q}_{0}$) and external (in ${\cal Q}_{1}$) states induces effective $2N$-, $3N$-, $\cdots$ $A$-body interactions in the subspace ${\cal Q}_{0}$ of localized states.
 
 The OQS solution in the subspace ${\cal Q}_0$ is found by solving equations:
 \begin{eqnarray}
{\cal H}(E)|\Psi_{\alpha}\rangle &=& {\cal E}_{\alpha}(E)|\Psi_{\alpha}\rangle \nonumber \\
\langle\Psi_{{\tilde \alpha}}|{\cal H}(E) &=& {\cal E}^*_{\alpha}\langle\Psi_{{\tilde \alpha}}|
\label{eq1}
\end{eqnarray}
 in the biorthogonal basis. The eigenstates $\Psi_{\alpha}$, $\Psi_{{\tilde \alpha}}$ satisfy: 
 $\langle\Psi_{{\tilde \alpha}}|\Psi_{\beta}\rangle = \delta_{\alpha\beta}$. For bound states, 
 ${\cal E}_{\alpha}(E)$ is real, whereas  physical resonances correspond to the poles of the scattering matrix. In general, all quantities calculated for unbound states are complex.
 
 The energy-dependent OQS solutions of Eqs.~(\ref{eq1})  are related to SM eigenvectors $\Phi_i$ in ${\cal Q}_{0}$ subspace by an orthogonal transformation
\begin{equation} 
\Psi_{\alpha}(E)=\sum_ib_{\alpha i}(E)\Phi_i
 \label{eq2}
\end{equation}
In a particular case of $^{14}$C discussed in this section, the decay-channel state is defined by the coupling of one neutron in the continuum of $^{14}$C  to $^{13}$C in a given SM state. 

Each decay threshold is associated with a non-analytic point of the scattering matrix. The coupling of different SM eigenfunctions to the same decay channel induces a mixing among the SM eigenfunction. Unitarity in a multichannel system implies that this external mixing of SM eigenfunctions changes whenever a new channel opens up.

The Hamiltonian of the SMEC consists of the  monopole-based SM interaction (referred to as WBP$-$ \cite{yuan16}) in the full psd model space plus the Wigner-Bartlett contact interaction for the coupling between SM states and the decay channels: 
$V_{12}=V_0 \left[ \alpha + \beta P^{\sigma}_{12} \right] \delta\langle\bf{r}_1-\bf{r}_2\rangle$, where $\alpha + \beta = 1$ and $P^{\sigma}_{12}$ is the spin exchange operator. The spin-exchange parameter $\alpha$ has a standard value of $\alpha = 0.73$. The radial single-particle wave functions (in ${\cal Q}_0$) and the scattering wave functions (in ${\cal Q}_1$) are generated by the Woods-Saxon potential which includes spin-orbit and Coulomb parts. The radius and diffuseness of the Woods-Saxon potential are 
$R_0=1.27 A^{1/3}$~fm and $a=0.67$~fm, respectively. The spin-orbit potential is $V_{\rm SO}=6.34$~MeV,  and the Coulomb part is calculated for a uniformly charged sphere with radius $R_0$. The depth of the central part for neutrons is adjusted to reproduce the measured neutron separation energy ($S_n$=8.176 MeV) for the  $p_{1/2}$ single-particle state. Similarly, the depth of the potential for protons is chosen to reproduce the measured proton separation energy  ($S_p$=20.831~MeV) for the  $p_{3/2}$ single-particle state. 
For each $J^{\pi}$ separately, the SM states are mixed via the coupling to 9 channels, including the elastic channel 
$[{^{13}}$C($1/2^-) \otimes {\rm p}(\ell_j)]^{J^{\pi}}$ and 8 inelastic channels $[{^{13}}$C($K^{\pi}) \otimes {\rm p}(\ell_j)]^{J^{\pi}}$ which correspond to the excited states of $^{13}$C: $K^{\pi}=1/2^+_1$, $3/2^-_1$, $5/2^+_1$, $5/2^+_2$, $3/2^+_1$, $7/2^+_1$, $5/2^-_1$, and $3/2^+_2$. 

\begin{figure}[h]
\begin{center}
\begin{minipage}{14pc}
\includegraphics[width=15pc]{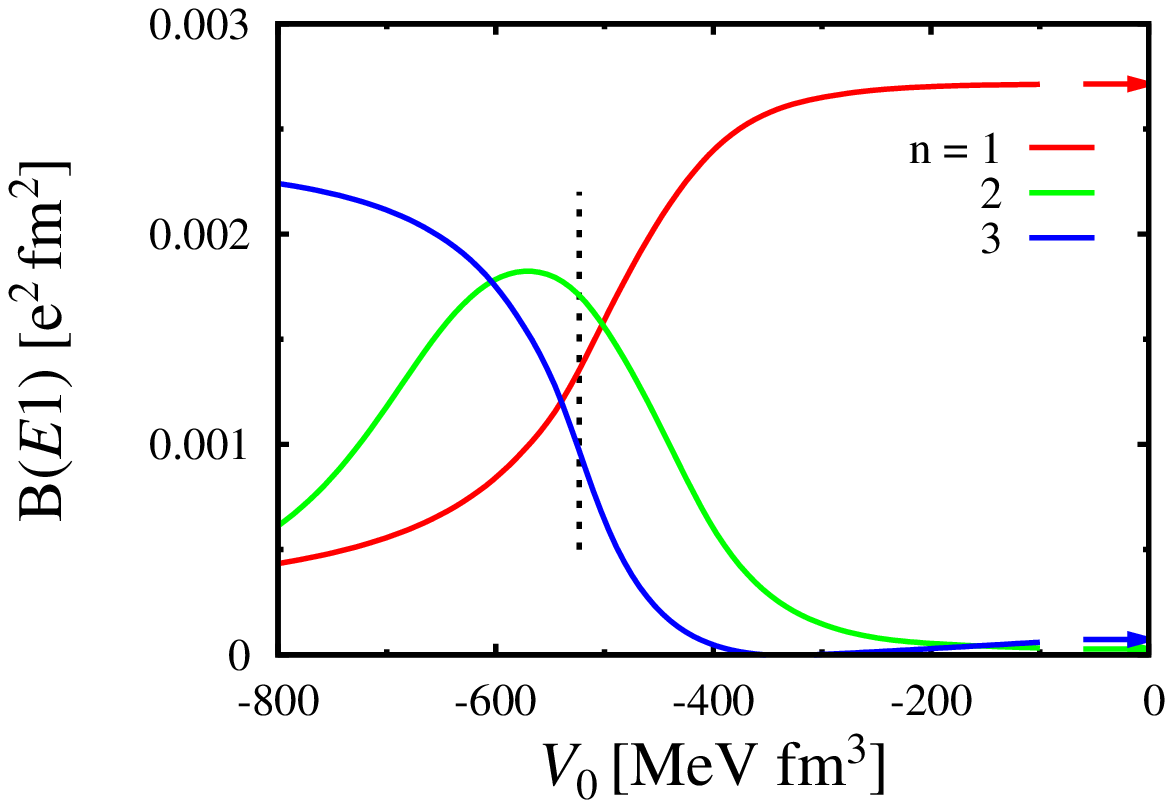}\hspace{3pc}%
\caption{\label{Fig1} (Color online) $B(E1)$ probabilities in SMEC for the $2^+_i \rightarrow 1^-_1$~ ($i = 1,2,3$) transitions of $^{14}$C as a function of the continuum-coupling constant. }
\end{minipage}\hspace{2pc}%
\begin{minipage}{15pc}
\includegraphics[width=15pc]{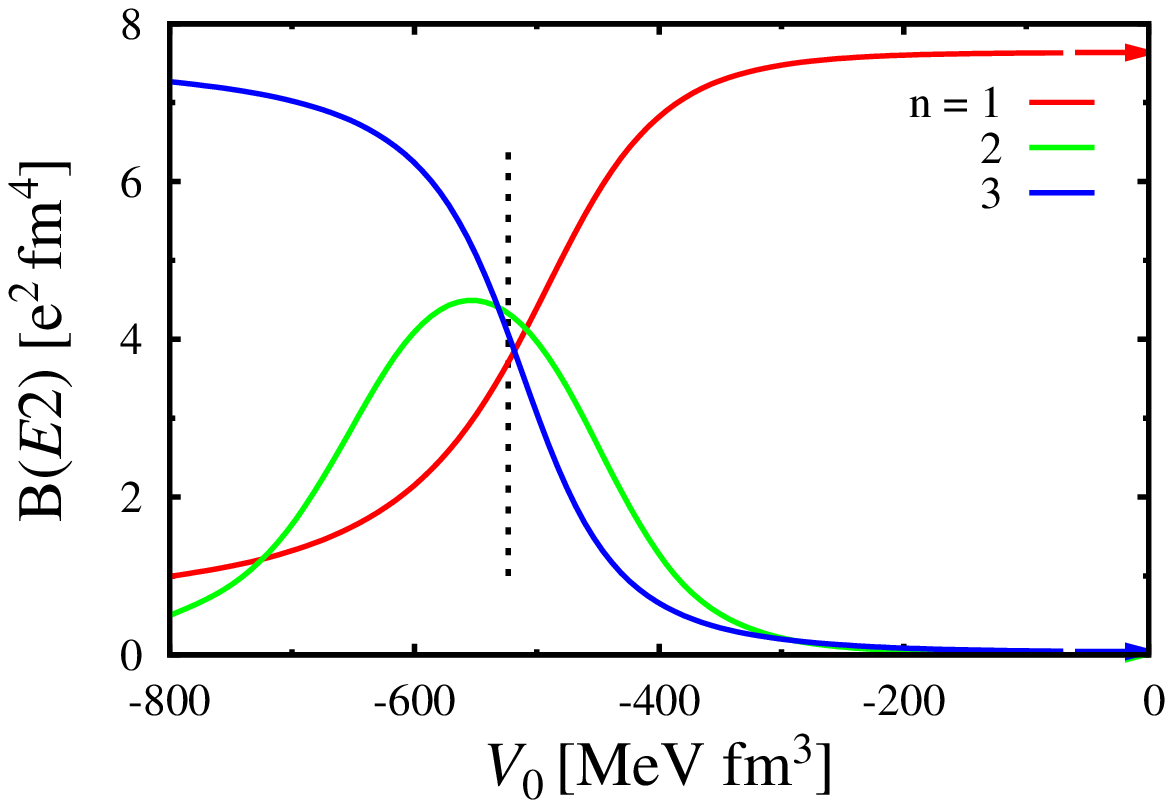}
\caption{\label{Fig2} (Color online) $B(E2)$ probabilities in SMEC for the $2^+_i \rightarrow 0^+_1$~ ($i = 1,2,3$) transitions of $^{14}$C as a function of the continuum-coupling constant.}
\end{minipage} 
\end{center}
\end{figure}

The near-threshold state $2^+_2$ at 8.318 MeV is located 142 keV above the one-neutron emission threshold 
$[{^{13}}$C($1/2^-) \otimes {\rm n}(p_{3/2})]^{2^+}$ and has a total width 3.4 keV \cite{ajz91}. Fig.~\ref{Fig1} displays the $B(E1)$ reduced transition probability for the $E1$ transition from the first three 2$^+$ excitations to the first 1$^-$ state, as a function of the coupling strength to the continuum $V_0$. For the transitions $2^+_i \rightarrow 1^-_1$~ ($i = 2,3$) involving resonance states, we show a real part of the reduced transition probability. In this figure, SM  results correspond to $V_0=0$. The dotted line in Fig.~\ref{Fig1} shows the value of the continuum-coupling strength $V_0 = -523.3$ MeV fm$^3$ for which the experimental $B(E2)$ probability for the $2^+_1 \rightarrow 0^+_1$ transition is reproduced in SMEC with WBP$-$ interaction. For this value of $V_0$, the $B(E1)$ probability for the $2^+_2 \rightarrow 1^-_1$ transition is enhanced by a factor $\sim 60$ and is the largest one among the considered $2^+_i \rightarrow 1^-_1$~ ($i = 1,2,3$) transitions. For higher $V_0$, $B(E1; 2^+_2 \rightarrow 1^-_1)$ becomes smaller and the decay from the third $2^+$ resonance dominates.

\begin{figure}[h]
\begin{center}
\begin{minipage}{14pc}
\includegraphics[width=15pc]{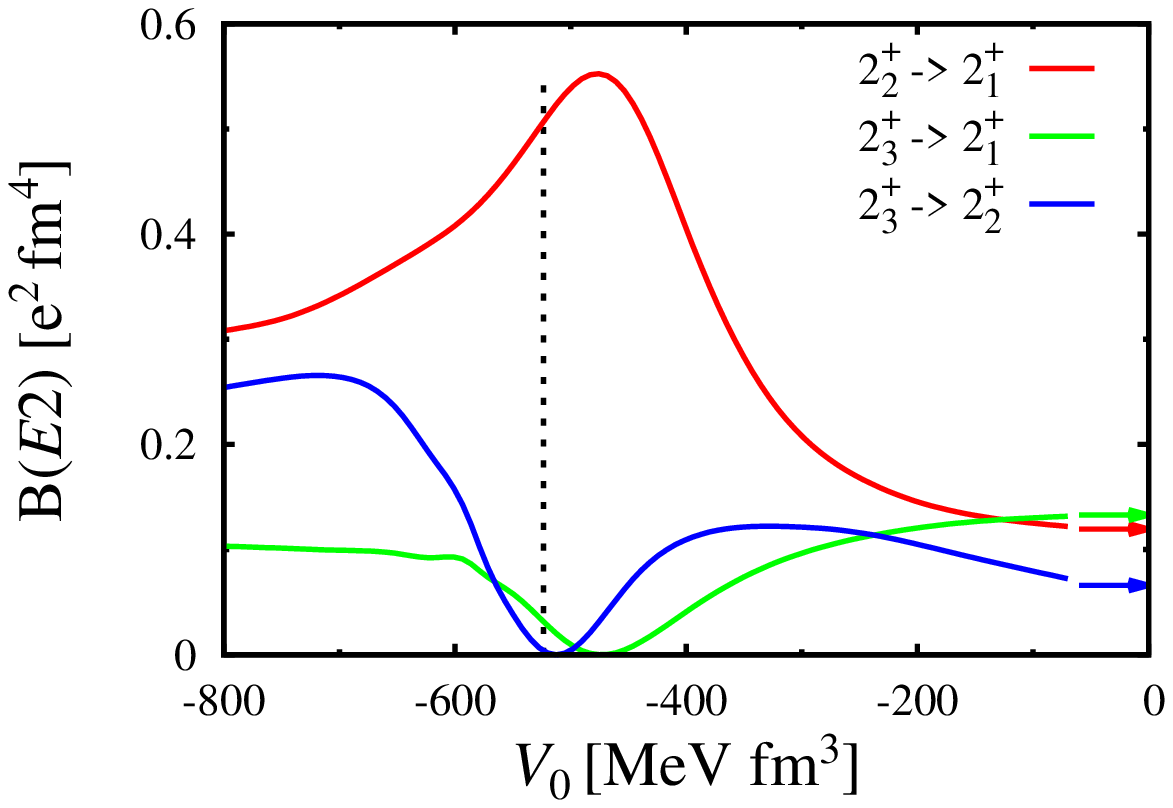}\hspace{3pc}%
\caption{\label{Fig3} (Color online) $B(E2)$ probabilities in SMEC for the $2^+_i \rightarrow 2^+_j$~ ($i,j = 1,2,3; i\neq j$) transitions of $^{14}$C as a function of the continuum-coupling constant.}
\end{minipage}\hspace{2pc}%
\begin{minipage}{14pc}
\includegraphics[width=15pc]{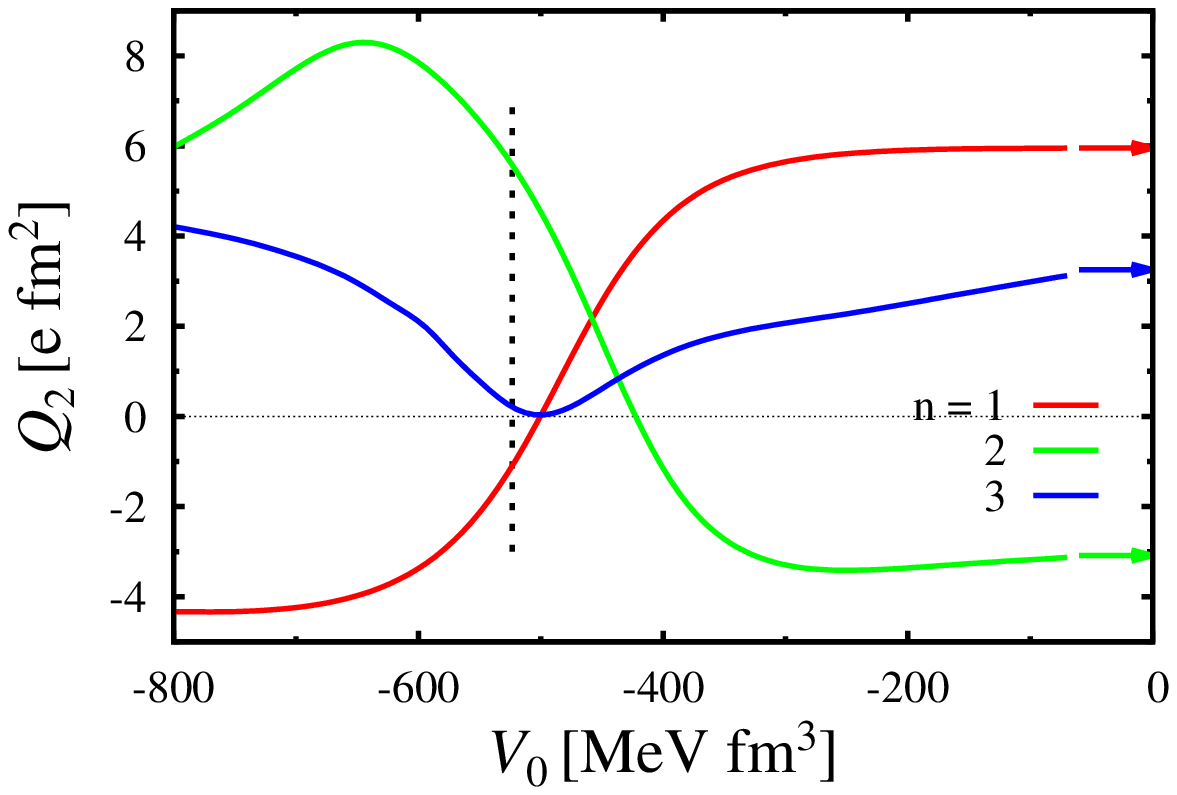}
\caption{\label{Fig4} (Color online) The spectroscopic quadrupole moment $Q_2$ for $2^+_i$~ ($i = 1,2,3$) as a function of the continuum-coupling constant.}
\end{minipage} 
\end{center}
\end{figure}
Fig.~\ref{Fig2} shows the $B(E2)$ reduced transition probability for the $E2$ transition from the 2$^+$ excitations to the ground state 0$^+$, as a function of the continuum coupling strength.  For the transitions $2^+_i \rightarrow 0^+_1$~ ($i = 2,3$), a real part of the reduced transition probability is shown. Changes of the $B(E2)$ reduced transition probabilities with $V_0$ resemble those shown in Fig.~\ref{Fig1} for the $B(E1)$ transition probabilities. Again, one can see a strong enhancement of the $B(E2)$ probability for the $2^+_2 \rightarrow 0^+_1$ transition which is enhanced by a factor of $\sim 340$ with respect to the SM value.

Fig.~\ref{Fig3} shows a real part of the $B(E2)$ reduced transition probability for the $E2$ transitions between the 2$^+_i$~ ($i = 1,2,3$) excitations, as a function of the continuum coupling strength. One can see that in a vicinity of $V_0 = -523.3$~MeV~fm$^3$, both the $2^+_3 \rightarrow 2^+_2$ and $2^+_3 \rightarrow 2^+_1$ transitions vanish. On the contrary, transition from the near-threshold $2^+_2$ excitation to the first $2^+_1$ state is strongly enhanced.
 
Fig.~\ref{Fig4} displays the dependence of the quadrupole moment for 2$^+_i$~ ($i = 1,2,3$) excitations on the continuum coupling strength. For resonances 2$^+_i$~ ($i = 2,3$), only a real part of the quadrupole moment is shown. One may see that $2^+_1$ and $2^+_2$ states exchange their nature in a vicinity of the value of the continuum-coupling strength $V_0 = -523.3$~MeV~fm$^3$. In this region of continuum-coupling strengths, the quadrupole moment for the $2^+_3$ excitation vanishes.

The interplay between Hermitian and anti-Hermitian parts of the effective Hamiltonian ${\cal H}(E)$ may lead to the coalescence of two eigenvalues, {\em i.e.}, to the formation of double poles of the scattering matrix, the so-called exceptional points (EPs)
\cite{rf:25}, which are the key ingredient of the configuration mixing mechanism in OQSs. Continuum induced mixing of SM eigenstates is strong if there are EPs or avoided crossings of SMEC eigenstates close to the real-$V_0$ axis (${\cal I}m(V_0) = 0$) in the complex-extended effective Hamiltonian \cite{oko2009}. EPs correspond to common roots of two equations:
\begin{equation}
 \frac{\partial^{(\nu)}}{\partial {\cal E}} {\rm det}\left[{\cal H}\left(E;V_0\right)  -{\cal E}I\right] = 0,~~~\nu=0,1.
\label{discr}
\end{equation}
Single-root solutions of Eq.~(\ref{discr}) correspond to EPs associated with either decaying or capturing states. Below the first decay threshold, half of all EPs have the asymptotic of a decaying state whereas the other half has the capturing state asymptotic. In the continuum, this symmetry is broken and the continuation in energy of a decaying pole may become a capturing pole, and {\em vice versa}. It should be noted that both decaying and capturing poles impact the configuration mixing in SMEC wave functions.

\begin{figure}
\begin{center}
\includegraphics{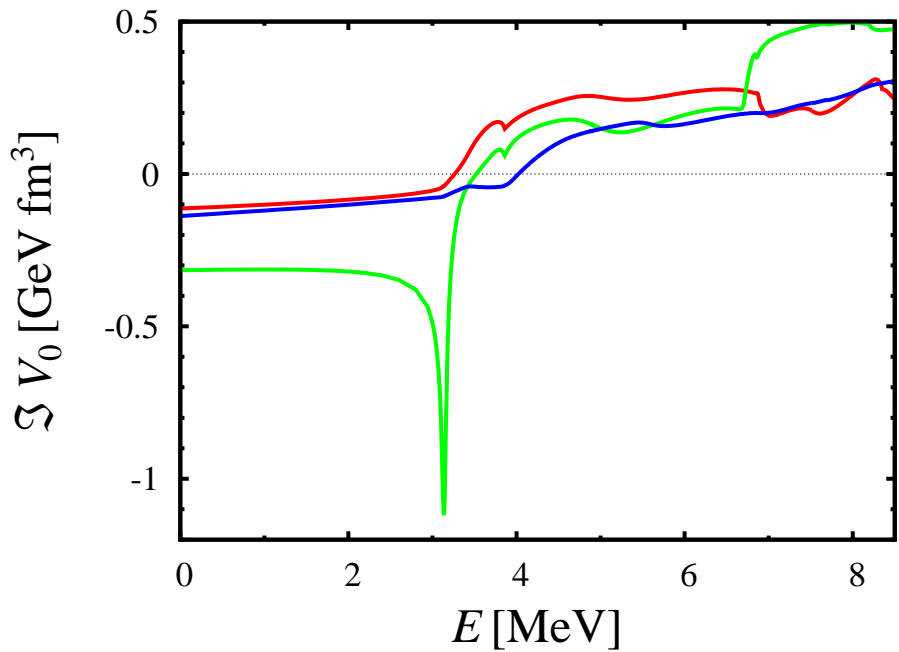}%
\end{center}
\caption{\label{Fig5} (Color online)  Trajectories of EPs for $2^+$ SMEC eigenstates are displayed in the ${\cal I}m(V_0)$ - $E$ plane. $E = 0$ corresponds to the lowest one-neutron emission threshold.}
\end{figure}
Fig.~\ref{Fig5} shows the 2D projection of the EP trajectories in the 3D space $[{\cal R}e(V_0), {\cal I}m(V_0), E]$. Three EP trajectories are crossing the real-$V_0$ axis (${\cal I}m(V_0) = 0$) in the narrow range of energies around $E \sim 3.5$ MeV. These trajectories are crucial for understanding the dependence of reduced transition probabilities $B(E1)$, $B(E2)$  and quadrupole moments on the continuum-coupling strength in 2$^+_1$ bound state and 2$^+_i$~ ($i = 2,3$) narrow resonances.

\section{Conclusions}

Shell model treatment of the OQSs allows the unification of nuclear structure and reactions. Mixing of shell model states via the continuum is at the origin of several generic, collective phenomena which can be studied in many mesosopic OQSs, such as atoms,  atomic nuclei, atomic clusters, quantum dots, quantum billiards, etc. In this respect, uniqueness and interest in atomic nucleus is due to the existence of nucleon in two states: neutron and proton, and their strong interaction. 

EPs strongly influence the spectrum and structure of low-energy resonances. Location of the EPs and the branch points, {\em i.e.}~the decay thresholds, do not vary in a systematic way from one nucleus to another. From one point of view this poses a tremendous challenge for the microscopic nuclear theory vis-a-vis the microscopic determination of effective nucleon-nucleon interaction.  From another point of view, with data that are sufficiently discriminatory, the continuum coupling constant can be fixed for a given nucleus. 

Near-threshold phenomena are the {\em terra incognita} of nuclear physics. Insight provided by the shell model for OQSs can help to define a new territory of nuclear spectroscopy studies. Results shown in Figs. \ref{Fig1}-\ref{Fig5} demonstrate a profound evolution of the nature of first three $2^+$ excitations in $^{14}$C as a function of the strength of the coupling between the SM 
$2^+$ eigenstates and 9 reaction channels, both open and closed ones. This evolution is modulated by the near-threshold $2^+_2$ resonance. We have shown that the near-threshold collectivization may have a noticeable effect on electromagnetic transitions and nuclear moments and, in particular, may modify the $\gamma$-decay selection rules for states close to the particle decay thresholds.

\section*{Acknowledgments}
Authors wish to thank R.J. Charity, B. Fornal, S. Leoni and L.G. Sobotka for their continuous encouragement and stimulating discussions. This work was supported by the COPIN and COPIGAL French-Polish scientific exchange programs.

\end{document}